\documentclass[oneside,reqno,12pt]{amsart}

\usepackage{rangecite}

\DeclareMathAlphabet{\mathscript}{OT1}{rsfs10}{m}{n}
\DeclareMathAlphabet{\mathscript}{OT1}{rsfs10}{m}{n}

\renewcommand{\mathcal}[1]{\mathscript{#1}}

\newcommand{\dud}[4]{{\smash{{{#1}_{#2}}^{#3}}}_{#4}}

\numberwithin{equation}{section}

\newcommand{\Begin}[2]{\begin{#1}\label{#2}}

\newcommand{\mfrac}[2] {\frac{\displaystyle #1}{\displaystyle #2}}
\newcommand{\half}{\textstyle \frac{1}{2} \displaystyle}

\newcommand{\al}{\alpha}
\newcommand{\be}{\beta}
\newcommand{\m}{\mu}
\newcommand{\n}{\nu}
\newcommand{\s}{\sigma}

\renewcommand{\sqrt}[1]{\left( #1 \right)^{1/2}}

\begin{document}

\vspace*{-0.575in}
\begin{flushleft}
\scriptsize{GR-QC/9406046, IC/91/209}
\end{flushleft}

\vskip 0.5in

\title[Energy-Momentum Complex in M\o ller's Theory]{Energy-Momentum Complex 
in M\o ller's \\ Tetrad Theory of Gravitation}
\author{F. I. Mikhail}
\address{Department of Mathematics \\ Faculty of Science \\ Ain Shams 
University \\ Cairo \\ Egypt}
\author{M. I. Wanas}
\address{Department of Astronomy \\ Faculty of Science \\ Cairo 
University \\ Cairo \\ Egypt}
\author{Ahmed Hindawi}
\address{International Centre for Theoretical Physics \\ Trieste \\ 
Italy}
\curraddr{Department of Physics \\ University of Pennsylvania \\ 
Philadelphia \\ PA~19104, USA}
\email{hindawi@frcu.eun.eg}
\author{E. I. Lashin}
\address{Department of Physics \\ Faculty of Science \\ Ain Shams
University \\ Cairo \\ Egypt}
\curraddr{International Centre for Theoretical Physics \\ Trieste \\ 
Itlay}
\email{lashin@frcu.eun.eg}
\thanks{Published in International Journal of Theoretical Physics 
{\textbf 32} (1993), 1627--1642}

\begin{abstract}

M\o ller's Tetrad Theory of Gravitation is examined with regard to the 
energy-momentum complex. The energy-momentum complex as well as the 
superpotential associated with M\o ller's theory are derived. M\o ller's 
field equations are solved in the case of spherical symmetry. Two 
different solutions, giving rise to the same metric, are obtained. The 
energy associated with one solution is found to be twice the energy 
associated with the other. Some suggestions to get out of this 
inconsistency are discussed at the end of the paper.

\vspace*{\baselineskip}


\end{abstract}

\maketitle

\thispagestyle{empty}

\renewcommand{\baselinestretch}{1.2}\large\normalsize

\section{Introduction}

The problem of defining an energy-momentum complex describing the energy 
contents of physical systems in General Relativity (GR) has been tackled 
by several authors 
\cite{Einstein-18,Bergmann-and-Thomson-53,Goldberg-58}. M\o ller 
\cite{Moller-58,Moller-61a,Moller-61b} pointed out that all expressions 
proposed previously for this quantity have some defects. He specified 
some properties to be satisfied. M\o ller \cite{Moller-61b} has shown 
that it is not possible to get an expression with these specifications 
using Riemannian space. Instead, he suggested using tetrad space. In 
fact, he was able to derive an expression for the energy-momentum
complex, possessing the properties mentioned before, in tetrad space.

The Lagrangian function from which the field equations of GR are derived 
is invariant under local tetrad rotation. Thus the field equations do 
not fix the field variables completely, leaving undefined six free 
functions. As a consequence, many different tetrad structures may give 
rise to the same metric specifying the gravitational field. And since 
the energy-momentum complex suggested by M\o ller is not invariant under 
local tetrad rotation, a certain metric, which is supposed to represent 
a single definite physical system, may be associated with more than one 
quantity expressing its energy and momentum contents. Thus the problem 
was not solved completely by the proposed expression mentioned above. 
M\o ller \cite{Moller-61b} suggested that the field equations of GR have 
to be modified in order not to allow such redundancy in solutions.

M\o ller \cite{Moller-78} modified GR by constructing a new field theory 
in the tetrad space. The field equations in this new theory were derived 
from a Lagrangian which is not invariant under local tetrad rotation. 
This theory has gained considerable attention 
\cite{Meyer-82,Saez-83,Saez-84,Saez-85,Saez-and-de-Juan-84}. The purpose 
of the present work is to examine this theory with regard to the energy-
momentum complex proposed by M\o ller \cite{Moller-61b}. In Section 2 we 
will review briefly M\o ller's tetrad theory of gravitation. The energy-
momentum complex associated with M\o ller's theory is derived in Section 
3. The structure of tetrad spaces with spherical symmetry is reviewed in 
Section~4. Two solutions of M\o ller's field equations are obtained in 
Section 5, using the tetrad of Section~4. A comparison between the two 
solutions is given in Section 6. In Section 7 the energy contents 
associated with each solution are evaluated. The results are discussed 
and concluded in Section 8.

\section{M\o ller's Tetrad Theory of Gravitation}

M\o ller \cite{Moller-78} constructed a gravitational theory using the 
tetrad space for its structure. His aim was to get a theory free from 
singularities while retaining the principle merits of GR as far as 
possible. In his theory the field variables are the 
16-tetrad components~${e_m}^\mu$. Hereafter we use Latin indices 
$(mn\ldots)$ for vector numbers and Greek indices $(\mu\nu\ldots)$ for 
vector components. All indices run from $0$ to $3$. The Riemannian 
metric is a derived quantity given by
\begin{equation}
g^{\mu\nu} := {e_m}^\mu {e_m}^\nu.
\label{2:11}
\end{equation}
We assume imaginary values for the vector ${e_0}^\mu$ in order for the 
above metric to have a Lorentz signature.

A central role in M\o ller's theory is played by the tensor
\begin{equation}
\gamma_{\mu\nu\sigma} := e_{m\mu} e_{m\nu;\sigma},
\end{equation}
where the semicolon denotes covariant differentiation using the 
Christoffel symbols. M\o ller \cite{Moller-78} considered the Lagrangian 
$L$ to be an invariant constructed from $\gamma_{\m\n\s}$ and 
$g_{\m\n}$. As he pointed out, the most simple possible independent 
expressions are
\Begin{equation}{2:3}
L^{(1)} := \Phi_\mu \Phi^\mu, \qquad 
L^{(2)} := \gamma_{\mu\nu\sigma}\gamma^{\mu\nu\sigma}, \qquad 
L^{(3)} := \gamma_{\mu\nu\sigma}\gamma^{\sigma\nu\mu},
\end{equation}
where $\Phi_\mu$ is the basic vector defined by
\begin{equation}
\Phi_\mu := {\gamma^\nu}_{\mu\nu}.
\end{equation}
These expressions $L^{(i)}$ in \eqref{2:3} are homogeneous quadratic 
functions in the first order derivatives of the tetrad field components.

M\o ller considered the simplest case, in which the Lagrangian $L$ is a 
linear combination of the quantities $L^{(i)}$, i.e., the Lagrangian 
density is given by
\Begin{equation}{2:4}
\mathcal{L}_{\text{M\o ller}} := (-g)^{1/2}\, (\al_1 L^{(1)} + \al_2 
L^{(2)} + \al_3 L^{(3)} ),
\end{equation}
where
\begin{equation}
g := \det(g_{\m\n}).
\end{equation}

Here, M\o ller chooses the constants $\al_i$ such that his theory gives 
the same results as GR in the linear approximation of weak fields. 
According to his calculations, one can easily see that if we choose
\begin{equation}
\al_1 = -1, \qquad \al_2=\lambda, \qquad \al_3=1-2\lambda,
\end{equation}
with $\lambda$ equals to a free dimensionless parameter of order unity, 
the theory will be in agreement with GR to the first order of 
approximation. For $\lambda=0$, M\o ller's field equations are identical 
with Einstein's equations
\Begin{equation}{2:8}
G_{\mu\nu} = -\kappa T_{\mu\nu}.
\end{equation}
For $\lambda\neq 0$, the field equations are given by
\Begin{align}{2:2}
G_{\mu\nu}+H_{\mu\nu} &= -\kappa T_{\mu\nu}, \\
F_{\mu\nu}  &= 0,
\label{2:13}
\end{align}
where
\Begin{equation}{2:10}
H_{\mu\nu} := \lambda \left[ \gamma_{\alpha\beta\mu} 
{\gamma^{\alpha\beta}}_{\nu} + \gamma_{\alpha\beta\mu} 
{\gamma_\nu}^{\alpha\beta} + \gamma_{\alpha\beta\nu} 
{\gamma_\mu}^{\alpha\beta} + g_{\mu\nu} \left( 
\gamma_{\alpha\beta\sigma}  \gamma^{\sigma\beta\alpha} - \half 
\gamma_{\alpha\beta\sigma}  \gamma^{\alpha\beta\sigma} \right) \right]
\end{equation}
and
\begin{equation}
F_{\mu\nu} := \lambda \left[ \Phi_{\mu,\nu} - \Phi_{\nu,\mu} - 
\Phi_{\alpha} \left( {\gamma^\alpha}_{\mu\nu} - {\gamma^\alpha}_{\nu\mu} 
\right) + \dud\gamma{\mu\nu}\alpha{;\alpha} \right].
\end{equation}
Equations \eqref{2:13} are independent of the free parameter $\lambda$. 
On the other hand the term $H_{\mu\nu}$ by which equations \eqref{2:2} 
deviate from Einstein's field equations \eqref{2:8} increases with 
$\lambda$, which can be taken of order unity without destroying the 
first order agreement with Einstein's theory in case of weak fields.

\section{Energy-Momentum Complex for M\o ller's Theory}

M\o ller \cite{Moller-61b} was able to find a general expression for an 
energy-momentum complex ${\mathcal{M}_\mu}^\nu$ that possesses all the 
required satisfactory properties, and formed its superpotential 
${\mathcal{U}_\mu}^{\nu\alpha}$ using the method of infinitesimal 
transformations:
\begin{equation}
{\mathcal{M}_\mu}^{\nu} := (-g)^{1/2}\, ( {T_\mu}^\nu + {t_\mu}^\nu ) = 
\dud{\mathcal{U}}\mu{\nu\alpha}{,\alpha},
\end{equation}
where
\begin{align}
(-g)^{1/2} \, {t_\mu}^\nu &:= \frac{1}{2\kappa} \left( \frac{\partial 
\mathcal{L}}{\partial { {e_m}^{\al} }_{,\nu}} \, { {e_m}^{\al} }_{,\mu} 
- {\delta_\mu}^\nu \mathcal{L} \right) \\
\intertext{and}
{\mathcal{U}_\mu}^{\nu\alpha} &:= \frac{1}{4\kappa} \left( 
\frac{\partial \mathcal{L}}{\partial { {e_m}^{\mu} }_{,\alpha}} \, 
{e_m}^{\nu} - \frac{\partial \mathcal{L}}{\partial { {e_m}^{\mu} 
}_{,\nu}} \, {e_m}^{\alpha} \right),
\label{3:1}
\end{align}
where $\mathcal{L}$ is the Lagrangian of the theory under consideration. 
For M\o ller's Lagrangian, as given by \eqref{2:4}, the superpotential 
\eqref{3:1} can be written in the form
\Begin{equation}{3:2}
{\mathcal{U}_\mu}^{\nu\sigma} = \frac{(-g)^{1/2}}{4\kappa} \left( \al_1 
{U_\mu}^{\nu\sigma} +\al_2 {V_\mu}^{\nu\sigma} + \al_3 
{W_\mu}^{\nu\sigma} \right),
\end{equation}
where ${U_\mu}^{\nu\sigma}, {V_\mu}^{\nu\sigma}, {W_\mu}^{\nu\sigma}$ 
correspond to $L^{(1)}, L^{(2)}, L^{(3)}$ respectively.

To evaluate the superpotential we have first (see Appendix A in 
\cite{Moller-61b})
\begin{equation}
\frac{\partial e_{m\mu;\nu}}{\partial \dud{e}{n}\rho{,\tau} } = -\half
{e_m}^{\alpha} \, {P_{\alpha\mu\nu\rho}}^{\sigma\tau} \, e_{n\sigma},
\end{equation}
where ${P_{\alpha\mu\nu}}^{\rho\sigma\tau}$ is a tensor of the form
\begin{equation}
{P_{\al\mu\nu}}^{\rho\s\tau} := {\delta_\al}^{\rho}
{g_{\mu\nu}}^{\s\tau} + {\delta_\mu}^{\rho} {g_{\nu\al}}^{\s\tau}
- {\delta_\nu}^{\s} {g_{\al\mu}}^{\s\tau}
\end{equation}
and ${g_{\mu\nu}}^{\s\tau}$ is the tensor
\begin{equation}
{g_{\mu\nu}}^{\s\tau} := {\delta_\mu}^{\s} {\delta_\nu}^{\tau}
- {\delta_\nu}^\s {\delta_\mu}^{\tau}.
\end{equation}
Thus we can get
\begin{align}
\frac{\partial L^{(1)}}{\partial {{e_m}^{\mu}}_{,\s}}  & = g^{\al\be}
\frac{\partial}{\partial {{e_m}^{\mu}}_{,\s}} \Phi_\al  \Phi_\be
\notag \\
& = 2 \Phi^\al \frac{\partial}{\partial {{e_m}^{\mu}}_{,\s}} \Phi_\al
\notag \\
& = 2 \Phi^\al {e_n}^\be \frac{\partial}{\partial 
{{e_m}^{\mu}}_{,\sigma}}
{e}_{n\al;\be} \notag \\
& = - \Phi^\al g^{\be\epsilon} {P_{\epsilon\al\be\mu}}^{\rho\s}
e_{m\rho}.
\end{align}
At last we get
\begin{align}
{U_\mu}^{\nu\sigma} & := \frac{\partial L^{(1)}}{\partial
{{e_m}^{\mu}}_{,\s}} {e_m}^{\nu} - \frac{\partial L^{(1)}}{\partial
{{e_m}^{\mu}}_{,\nu}} {e_m}^{\s} \notag \\
& = -2 \Phi^\al g^{\be\epsilon} g_{\m\tau}
{P_{\epsilon\al\be}}^{\tau\n\s}.
\label{3:3}
\end{align}
Similarly we can write
\begin{equation}
\begin{split}
{V_\mu}^{\nu\sigma} & = -2 \gamma^{\epsilon\al\be} g_{\m\tau}
{P_{\epsilon\al\be}}^{\tau\n\s}, \\
{W_\mu}^{\nu\sigma} & = -2 \gamma^{\be\al\epsilon} g_{\m\tau}
{P_{\epsilon\al\be}}^{\tau\n\s}.
\end{split}
\label{3:5}
\end{equation}

The final expression for the superpotential for M\o ller's Theory can be 
obtained by substituting from \eqref{3:3} and \eqref{3:5} 
and using the values of the parameters $\al_1, \al_2, \al_3$ given in 
Section~2 in \eqref{3:2}, to get
\Begin{equation}{3:6}
{\mathcal{U}_\mu}^{\nu\s} = \frac{\sqrt{-g}}{2\kappa}
{P_{\epsilon\al\be}}^{\tau\nu\sigma} \left( \Phi^\al g^{\be\epsilon}
g_{\mu\tau} -\lambda g_{\tau\mu}
\gamma^{\epsilon\al\be} - (1-2\lambda) g_{\tau\mu}
\gamma^{\be\al\epsilon} \right).
\end{equation}

\section{Spherically Symmetric Tetrad Spaces}

The structure of tetrad spaces with spherical symmetry has been studied
by Robertson \cite{Robertson-32}. The four tetrad vectors defining such
structure, as given by Robertson, can be written as
\begin{equation}
\begin{split}
{e_0}^0 &= A, \qquad {e_0}^a = D X^a, \qquad {e_a}^0 = E X^a, \\
{e_a}^b &= F X^a X^b + {\delta_a}^b B + \epsilon_{abc}SX^c,
\end{split}
\label{4:1}
\end{equation}

where $A$, $B$, $D$, $E$, $F$, $S$ are functions of 
$r=\left(\sum_{a=1}^3 X^aX^a\right)^{1/2}$ and $a$, $b$, $c$ run 
from~$1$~to~$3$.

Robertson has shown that:
\begin{itemize}
\item[1.] Improper rotations are admitted if and only if $S=0$. In this
case the tetrad \eqref{4:1} takes the form
\begin{equation}
\begin{split}
{e_0}^0 & = A, \qquad {e_0}^a = D X^a, \qquad {e_a}^0 = E X^a, \\
{e_a}^b & = F X^a X^b + {\delta_a}^b B.
\end{split}
\label{4:2}
\end{equation}
\item[2.] The functions $E$ and $F$ can be eliminated by mere coordinate
transformations, leaving the tetrad in the simpler form
\begin{equation}
{e_0}^0 = A, \qquad {e_0}^a = D X^a, \qquad {e_a}^b = {\delta_a}^b B.
\label{4:3}
\end{equation}
\end{itemize}

Three important remarks are reported here:
\begin{itemize}
\item[1.] The tetrad used by M\o ller \cite{Moller-78} in application of
his theory, is a special case of the above tetrad \eqref{4:3}, in which
the function $D$ is taken to be zero. Thus one may expect to obtain more
solutions when using the more general tetrad \eqref{4:3}.
\item[2.] Since one has to take the vector ${e_0}^\mu$ to be imaginary, 
in order to ensure the Lorentz signature of the metric, the functions 
$A$ and $D$ have to be taken to be imaginary. 
\item[3.] It is more convenient, for the sake of computations, to use
the tetrad \eqref{4:3} in spherical polar coordinates where it takes the
form
\Begin{equation}{4:4}
{e_m}^ \mu=
\begin{pmatrix}
A \rule{0ex}{3.2ex} & Dr & 0 & 0 \\
0 \rule{0ex}{3.2ex} & B\sin\theta\cos\phi &
\mfrac{B}{r}\cos\theta\cos\phi
  & -\mfrac{B}{r}\mfrac{\sin\phi}{\sin\theta} \\
0 \rule{0ex}{3.2ex} & B\sin\theta\sin\phi &
\mfrac{B}{r}\cos\theta\sin\phi
  & \mfrac{B}{r}\mfrac{\cos\phi}{\sin\theta} \\
0 \rule{0ex}{3.2ex} & B\cos\theta & -\mfrac{B}{r}\sin\theta & 0
\end{pmatrix}.
\end{equation}
\end{itemize}

\section{Solutions of M\o ller's Field Equations}

Using the tetrad \eqref{4:4} to solve M\o ller's field equations
\eqref{2:2} and \eqref{2:13} we find that equation \eqref{2:13} is
satisfied identically, and also that $H_{\m\n}$ as given by \eqref{2:10}
vanishes identically. Thus for spherically symmetric exterior solutions,
M\o ller's field equations are reduced to Einstein's field equations of
GR, namely
\Begin{equation}{5:1}
G_{\mu\nu} = 0.
\end{equation}
Einstein tensor $G_{\m\n}$ may be evaluated using the Riemannian metric
derived from \eqref{4:4} via the relation \eqref{2:11}. It is easy to
get
\begin{equation}
\begin{aligned}
g_{00} &= \mfrac{D^2r^2+B^2}{A^2B^2}, \\
g_{22} &= \mfrac{r^2}{B^2},
\end{aligned}
\qquad
\begin{aligned}
g_{01} & = g_{10} = -
\frac{Dr}{AB^2}, \qquad  g_{11} = \frac{1}{B^2}, \\
g_{33} & = \frac{r^2\sin^2\theta}{B^2}.
\end{aligned}
\label{5:2}
\end{equation}
The corresponding field equations \eqref{5:1} give rise to the following
set of differential equations:
\begin{multline}
2r^3BDB'D' -2r^2B^2DD' +8r^2BD^2B' - 5r^3D^2{B'}^2 + 2rB^3B'' \\
+ 2r^3BD^2B'' - 3rB^2D^2 + 4B^3B' - 3 r B^2{B'}^2 =0,
\end{multline}
\begin{multline}
5r^3D^2{B'}^2 -2r^3BD^2B'' - 2r^3BDB'D' + 2r^2B^2DD' - 8r^2BD^2B' \\
- 2rB^3B'' + 3rB^2D^2 + 3rB^2{B'}^2 - 4B^3B' = 0,
\end{multline}
\begin{multline}
2r^2AB^2DD'  -2r^3ABD^2B'' - 2r^3ABDB'D' - 2AB^3B' + 5r^3AD^2{B'}^2 \\
+ 3rAB^2D^2 - 8r^2ABD^2B' + rAB^2{B'}^2 + 2rB^3A'B' - 2B^4A'  = 0,
\end{multline}
\begin{multline}
r^3A^2B^2DD''-2r^3A^2BD^2B''+r^3A^2B^2{D'}^2-5r^3A^2BDB'D'-AB^4A' \\
+5r^3A^2D^2{B'}^2 -r^3AB^2D^2A'' -3r^3AB^2DA'D'-A^2B^3B'-rAB^4A'' \\
+rA^2B^2{B'}^2 +3rA^2B^2D^2+2rB^4{A'}^2-rA^2B^3B''+3r^3ABD^2A'B' \\
-8r^2A^2BD^2B' +2r^3B^2D^2{A'}^2  +6r^2A^2B^2DD'-4r^2AB^2D^2A' = 0,
\end{multline}
where the primes refer to differentiation with respect to $r$.

The trivial flat space-time solution for such equations is obtained by
taking
\begin{equation}
A = i, \qquad B = 1, \qquad D = 0.
\end{equation}
A first non-trivial solution can be obtained by taking $D=0$ and solving
for $A$ and $B$. In fact, this is the case studied by M\o ller
\cite{Moller-78}, where he obtained the solution
\begin{alignat}{3}
A & = i\frac{(1+m/2r)}{(1-m/2r)}, & \qquad B & = (1+m/2r)^{-2}. \\
\intertext{Hence, we get from \eqref{4:3} directly the tetrad (in
cartesian coordinates):}
{e_0}^0 & = i\mfrac{(1+m/2r)}{(1-m/2r)}, & \qquad {e_a}^a & = 
(1+m/2r)^{-2},
\label{5:12}
\end{alignat}
with the associated Riemannian metric
\Begin{equation}{5:20}
ds^2 = - \frac{(1-m/2r)^2}{(1+m/2r)^2}dt^2+(1+m/2r)^4(dX^2+dY^2+dZ^2),
\end{equation}
i.e., Schwarzschild metric in its isotropic form.

A second non-trivial solution can be obtained by taking $A=1$, $B=1$, $D
\neq 0$ and solving for $D$. In this case the resulting field equations
can be integrated directly to give
\Begin{equation}{5:10}
D=i\sqrt{\frac{2m}{r^3}}.
\end{equation}
Hence, we get from \eqref{4:3} the following tetrad (in cartesian
coordinates):
\Begin{equation}{5:16}
{e_0}^0 = i, \qquad {e_0}^a  = i\sqrt{\mfrac{2m}{r^3}}X^a, \qquad
{e_a}^{b}
= {\delta_a}^b.
\end{equation}
The metric associated with the above tetrad is
\begin{equation}
ds^2 = - (1-2m/r)dt^2  + \sum_{a=1}^3 dX^a dX^a - 2\sqrt{\frac{2m}{r^3}}
X^a dt\, dX^a.
\end{equation}
A simpler form for the above metric can be obtained if it is
written in polar coordinates. Substituting directly in \eqref{4:4} for
the value of $D$ as given by \eqref{5:10}, we get the tetrad (in polar
coordinates)
\begin{equation}
{e_m}^\mu=
\begin{pmatrix}
i \rule{0ex}{3.2ex} & i\left(\dfrac{2m}{r} \right)^{1/2} & 0 & 0 \\
0 \rule{0ex}{3.2ex} & \sin\theta\cos\phi & \mfrac{\cos\theta\cos\phi}{r}
  & - \mfrac{\sin\phi}{r\sin\theta} \\
0 \rule{0ex}{3.2ex} & \sin\theta\sin\phi & \mfrac{\cos\theta\sin\phi}{r}
  & \mfrac{\cos\phi}{r\sin\theta} \\
0 \rule{0ex}{3.2ex} & \cos\theta & -\mfrac{\sin\theta}{r} & 0
\end{pmatrix},
\end{equation}
with the associated Riemannian metric
\Begin{equation}{5:11}
ds^2=-(1-2m/r)dt^2- 2 \sqrt{\frac{2m}{r}}\,dtdr+dr^2+r^2d\theta^2 +
r^2\sin^2\theta d\phi^2.
\end{equation}
It is to be noted here that $m$, in the above metric \eqref{5:11}, is a
mere constant of integration. It will be shown in the next section that
$m$, indeed plays the role of the mass producing the field, and thus
justifies the use of its name.

\section{Comparison of the Two Solutions}

Our aim in this section is to compare the two different solutions
obtained in Section~5, for M\o ller's field equations. The first step is
to eliminate the cross term appearing in the metric \eqref{5:11} of the
second solution. This can be easily done by performing the coordinate
transformation
\Begin{equation}{6:1}
t \rightarrow t + \int \frac{-iDr}{(1-D^2r^2)} dr
\end{equation}
and keeping the spatial coordinates unchanged. One can get the
transformed tetrad in the form
\begin{alignat}{2}
{e_0}^0 & = i\mfrac{1}{(1-2m/r)}, & \qquad {e_0}^1 & =
i\sqrt{\mfrac{2m}{r}}, \notag \\
{e_1}^0 & = \sqrt{\mfrac{2m}{r}} \mfrac{\cos\phi\sin\theta}{1-2m/r}, &
\qquad {e_1}^1 & = \cos\phi\sin\theta, \notag \\
{e_1}^2 & = \mfrac{\cos\phi\cos\theta}{r}, & \qquad {e_1}^3  & = -
\mfrac{\sin\phi}{r\sin\theta}, \label{6:2} \\
{e_2}^0 & =  \sqrt{\mfrac{2m}{r}} \mfrac{\sin\phi\sin\theta}{1-2m/r}, &
\qquad {e_2}^1  & =  \sin\phi\sin\theta, \notag \\
{e_2}^2 & = \mfrac{\cos\theta\sin\phi}{r}, & \qquad {e_2}^3  & =
\sqrt{\mfrac{2m}{r}}\mfrac{\cos\theta}{1-2m/r}, \notag \\
{e_3}^1 & = \cos\theta, & \qquad {e_3}^2 & = -\mfrac{\sin\theta}{r}. 
\notag
\end{alignat}
The metric associated with the above tetrad can be computed either
directly from the tetrad, or by applying the same coordinate
transformation \eqref{6:1} to the metric \eqref{5:11}. In both cases we
get
\begin{equation}
ds^2=-(1-2m/r)dt^2+(1-2m/r)^{-1}dr^2+r^2d\theta^2+r^2\sin^2\theta
d\phi^2,
\end{equation}
i.e., Schwarzschild metric in its standard form, in which $m$, the
constant of integration, plays the role of the mass of the source of the
field.

Now to be able to compare the two solutions \eqref{5:12} and
\eqref{6:2}, we transform the second one \eqref{6:2} to a coordinate
system such that its Riemannian metric takes the isotropic form in
cartesian coordinates, i.e., the same coordinates of the first solution
\eqref{6:2}. The first coordinate transformation
(cf.\ \cite{Eddington-21} p.\ 93) is
\begin{equation}
r \rightarrow r\left(1+\frac{m}{2r}\right)^2.
\label{6:3}
\end{equation}
Applying this coordinate transformation yields the following tetrad
\begin{alignat}{2}
{e_0}^0 & = i\mfrac{(1+m/2r)^2}{(1-m/2r)^2}, & \qquad {e_0}^1 & =
i\mfrac{2 \sqrt{m/2r}}{(1-m/2r)(1+m/2r)^2}, \notag \\
{e_1}^0 & = 2 \sqrt{\mfrac{m}{2r}} \mfrac{(1+m/2r)\cos\phi\sin\theta}
{(1-m/2r)^2}, & \qquad {e_1}^1 & = \mfrac{\cos\phi\sin\theta} {(1-m/2r)
(1+m/2r)}, \notag \\
{e_1}^2 & = \mfrac{\cos\phi\cos\theta}{r(1+m/2r)^2}, & \qquad {e_1}^3  &
=  -\mfrac{\sin\phi}{r\sin\theta(1-m/2r)^2}, \notag \\
{e_2}^0 & = 2\sqrt{\mfrac{m}{2r}}\mfrac{(1+m/2r)\sin\phi\sin\theta} {(1-
m/2r)^2}, & \qquad {e_2}^1 & = \mfrac{\sin\phi\sin\theta}{(1+m/2r) (1 -
m/2r)}, \label{6:4} \\
{e_2}^2 & = \mfrac{\cos\theta\sin\phi}{r(1+m/2r)^2}, & \qquad
{e_2}^3 & = \mfrac{\cos\phi}{r\sin\theta(1+m/2r)^2}, \notag \\
{e_3}^0 & = 2\sqrt{\mfrac{m}{2r}}\mfrac{(1+m/2r)\cos\theta}{(1-m/2r)^2},
& \qquad {e_3}^1 & = \mfrac{\cos\theta}{(1+m/2r)(1-m/2r)}, \notag \\
{e_3}^3 & = -\mfrac{\sin\theta}{r(1+m/2r)^2}. \notag
\end{alignat}
The metric associated with the above tetrad \eqref{6:4} is
\Begin{equation}{6:5}
ds^2 = -\frac{(1-m/2r)^2}{(1+m/2r)^2} dt^2 + (1+m/2r)^4 (dr^2 +
r^2d\theta^2+r^2\sin^2\theta d\phi^2).
\end{equation}

The last step is to transform the tetrad (\ref{6:4}), along with its
metric \eqref{6:5}, into cartesian coordinates,
\begin{alignat}{2}
{e_0}^0 & = i\mfrac{(1+m/2r)^2}{(1-m/2r)^2}, & \qquad {e_0}^1 & =
i\mfrac{2 \sqrt{m/2r}X}{r(1-m/2r)(1+m/2r)^2}, \notag \\
{e_0}^2 & =  i\mfrac{2 \sqrt{m/2r}Y}{r(1-m/2r)(1+m/2r)^2}, & \qquad
{e_0}^3 & =  i\mfrac{2 \sqrt{m/2r}Z}{r(1-m/2r)(1+m/2r)^2}, \notag \\
{e_1}^0 & = \mfrac{2\sqrt{2m/r}(1+m/2r)X}{(1-m/2r)^2r}, & \qquad {e_1}^1
&
= \mfrac{(1-m/2r)r^2+2X^2(m/2r)}{(1+m/2r)^2(1-m/2r)r^2}, \notag \\
{e_1}^2 & = \mfrac{2XY(m/2r)}{(1+m/2r)^2(1-m/2r)r^2}, & \qquad {e_1}^3 &
=
\mfrac{2XZ(m/2r)}{(1+m/2r)^2(1-m/2r)r^2}, \label{6:10} \\
{e_2}^0 & = \mfrac{2\sqrt{2m/r}(1+m/2r)Y}{(1-m/2r)^2r}, & \qquad {e_2}^1
&
= \mfrac{2XY(m/2r)}{(1+m/2r)^2(1-m/2r)r^2}, \notag \\
{e_2}^2 & = \mfrac{(1-m/2r)r^2+2Y^2(m/2r)}{(1+m/2r)^2(1-m/2r)r^2}, &
\qquad
{e_2}^3 & = \mfrac{2YZ(m/2r)}{(1+m/2r)^2(1-m/2r)r^2}, \notag
\end{alignat}
\begin{alignat}{2}
{e_3}^0 & = \mfrac{2\sqrt{2m/r}(1+m/2r)Z}{(1-m/2r)^2r}, & \qquad {e_3}^1
&= \mfrac{2XZ(m/2r)}{(1+m/2r)^2(1-m/2r)r^2}, \notag \\
{e_3}^2 & = \mfrac{2YZ(m/2r)}{(1+m/2r)^2(1-m/2r)r^2}, & \qquad {e_3}^3 &
=\mfrac{(1-m/2r)r^2+2Z^2(m/2r)}{(1+m/2r)^2(1-m/2r)r^2}. \notag
\end{alignat}
The metric derived from the above tetrad \eqref{6:10} is now identical
with the metric derived from the first solution \eqref{5:20}, namely
\begin{equation}
ds^2 = -\frac{(1-m/2r)^2}{(1+m/2r)^2} dt^2+(1+m/2r)^4 (dX^2+dY^2+dZ^2).
\end{equation}

Thus we have two exact solutions of M\o ller field equations, each of
which leads to the same metric, the Schwarzschild metric in its
isotropic form in cartesian coordinates. We notice that the second
solution \eqref{6:10} is of the form of the original Robertson tetrad
\eqref{4:2}. This should be expected, since the coordinate
transformations we have performed on the second solution reproduce the
functions $E$ and $F$, eliminated before by coordinate transformation.
Hence, we can put these two solutions into a concise form, as shown in
Table I.

\begin{center}
\vspace*{1mm}
\begin{tabular}{c}
{\sc Table I.} Comparison of the Two Solutions \\[1mm]
\begin{tabular}{ccc} \hline \hline
Function $\rule{0ex}{2.5ex}$ & First Solution (5.9) & Second Solution 
(6.7)
\\[1mm] \hline
$A$ & $(1+m/2r)(1-m/2r)^{-1} \rule{0ex}{3ex}$ & $(1+m/2r)^2(1-m/2r)^{-
2}$
\\[3mm]
$B$ & $(1+m/2r)^{-2}$ & $(1+m/2r)^{-2}$
\\[3mm]
$D$ & $0$ & $\mfrac{2}{r}\sqrt{\displaystyle \mfrac{m}{2r}}
(1+m/2r)^{-2}(1-m/2r)^{-2}$ \\[3mm]
$E$ & $0$ & $\mfrac{2}{r}\sqrt{\displaystyle \mfrac{m}{2r}}
(1+m/2r)(1-m/2r)^{-2}$ \\[3mm]
$F$ & $0$ & $\mfrac{2}{r} {\displaystyle \left( \mfrac{m}{2r} \right)}
(1+m/2r)^{-2}(1-m/2r)^{-1}$ \\[3mm] \hline \hline
\end{tabular}
\\
\end{tabular}
\end{center}

\vspace*{0.5\baselineskip}

The important result obtained in this section is that we have been able
to derive two different solutions for M\o ller's field equations, the
Riemannian metrics associated with these two solutions are identical,
namely the Schwarzschild metric in its isotropic form. Since M\o ller's
theory is a pure gravitational theory, the above two solutions have to
be equivalent in the sense that they describe the same physical
situation, viz a static spherically symmetric gravitational field with a
source of mass $m$. In what follows we examine the equivalence of these
solutions by calculating the energy associated with each of them, using
the superpotential derived for M\o ller's theory in Section~3.

\section{The Energy Associated with each Solution}

Now we use the superpotential of M\o ller's theory derived in Section~3
to evaluate the energy associated with each of the two solutions given
in the previous table. The components of the superpotential that
contribute to the total energy are ${\mathcal{U}_0}^{0\sigma}$ only. 
Thus substituting from the first solution \eqref{5:12} into \eqref{3:6}, 
we get the following non-vanishing values:
\Begin{equation}{7:1}
{\mathcal{U}_0}^{0a} = \mfrac{4X^a}{\kappa r^2}\mfrac{m}{2r} (1-m/2r).
\end{equation}
The total energy is given by (cf.\ \cite{Moller-58})
\Begin{equation}{7:2}
E = \lim_{r\rightarrow\infty} \int_{r=\text{const}} {\mathcal{U}_0}^{0a} 
n_a dS,
\end{equation}
where $n_a$ is a unit 3-vector normal to the surface element $dS$.
Substituting from \eqref{7:1} into \eqref{7:2}, we get
\begin{equation}
E = \frac{8 \pi m}{\kappa} = m.
\end{equation}
In non-relativistic units, the above result appears as the mass of the
source times the square of the speed of light. This is a very
satisfactory result, and it should be expected.

Now let us turn our attention to the second tetrad \eqref{6:10}.
Computing the required components of the superpotential, we get
\begin{equation}
{\mathcal{U}_0}^{0a} = \mfrac{8X^a}{\kappa r^2}\mfrac{m}{2r}.
\end{equation}
These lead to a total energy
\begin{equation}
E=2m.
\end{equation}
That is twice the gravitational mass!

\section{Discussion and Conclusion}

The energy-momentum complex for M\o ller's tetrad theory of gravitation
is derived, using M\o ller's Lagrangian. Two different exact solutions
of M\o ller's field equations are obtained for the case of spherical
symmetry. The energy content of each solution is evaluated using the
derived superpotential. It is shown that, although the two solutions
give rise to the same Riemannian metric (the Schwarzschild metric), they
give two different values for the energy content. This shows a certain
type of inconsistency in M\o ller's theory.

The following suggestions may be considered to get out of this
inconsistency:
\begin{itemize}
\item[1.] The energy-momentum complex suggested by M\o ller
\cite{Moller-61b} is not quite adequate, though it has the most
satisfactory properties.
\item[2.] Many authors believe that a tetrad theory should describe more
than a pure gravitational field. In fact, M\o ller himself
\cite{Moller-61b} considered this possibility in his earlier trials to
modify GR. In these theories, the most successful candidates for the
description of the other physical phenomenon are the skew-symmetric
tensors of the tetrad space, e.g., $\Phi_{\mu;\nu}-\Phi_{\nu;\mu}$. The 
most striking remark here is that: All the skew-symmetric tensors vanish 
for the first solution, but not all of them do so for the second one. 
Some authors, e.g., \cite{Einstein-30,Mikhail-and-Wanas-77}, believe 
that these tensors are related to the presence of an electromagnetic 
field. Others, e.g., \cite{Muller-Hoissen-and-Nitsch-83}, believe that 
these tensors are closely connected to the spin phenomenon. There are a 
lot of difficulties to claim that M\o ller's theory deserves such wider
interpretation. This needs a lot of investigations before arriving at a
concrete conclusion.
\item[3.] The last possibility is that M\o ller's theory is in need to
be generalized rather than to be reinterpreted. There are already some
generalizations of M\o ller's theory. M\o ller himself considered this
possibility at the end of his 1978 paper \cite{Moller-78}, by including
terms in the Lagrangian other than the simple quadratic terms
$L^{(i)}$'s. S\'aez \cite{Saez-83} has generalized M\o ller's theory
in a very elegant and natural way into scalar tetradic theories of
gravitation. In these theories the question is: Do the field equations
fix the tetradic geometry in the case of spherical symmetry? This
question was discussed in length by S\'aez \cite{Saez-86}. The results
of the present paper can be considered as a first step to get a
satisfactory answer to this question. In 1982, Meyer \cite{Meyer-82} has
shown that M\o ller's theory is a special case of Poincar\'e Gauge
Theory constructed by Hehl et al.\ \cite{Hehl-et-al-80}. Thus
Poincar\'e Gauge Theory can be considered as another satisfactory
generalization of M\o ller's theory.
\end{itemize}

\section*{Acknowledgments}
The authors would like to express their gratitude to Dr.\ M. Melek,
Cairo University, for his stimulating discussions. They would also like
to thank  Professor M. M. Shalaby, Ain Shams University, for providing
REDUCE 3.3 that has been used during carrying out this work. One of the
authors (A. H.) is extremely grateful to Professor Adnan Hamoui, ICTP,
for his comments and suggestions. He would also like to thank Professor
Abdus Salam, the International Atomic Energy Agency and UNESCO for
hospitality at the International Centre for Theoretical Physics,
Trieste.

\ifx\undefined\bysame
\newcommand{\bysame}{\leavevmode\hbox to3em{\hrulefill}\,}
\fi

\end{document}